\documentclass[prl,twocolumn,superscriptaddress,floatfix,showpacs,preprintnumbers]{revtex4}

\usepackage[latin2]{inputenc}
\usepackage{amsmath}
\usepackage{amssymb}
\usepackage{epsfig}
\usepackage{bm}
\usepackage{color}

\begin{document}

\title{Nonadditivity of Fluctuation-Induced Forces 
in Fluidized Granular Media}

\author{M.\ Reza Shaebani}
\email{reza.shaebani@uni-due.de}
\affiliation{Department of Theoretical Physics, University of
Duisburg-Essen, 47048 Duisburg, Germany}
\author{Jalal Sarabadani}
\altaffiliation{present address: Max Planck Institute
for Polymer Research, D-55128 Mainz, Germany.}
\affiliation{Department of Physics, University of Isfahan,
Isfahan 81746, Iran} 
\author{Dietrich E.\ Wolf}
\affiliation{Department of Theoretical Physics, University of
Duisburg-Essen, 47048 Duisburg, Germany}

\date{\today}

\begin{abstract}
We investigate the effective long-range interactions between
intruder particles immersed in a randomly driven granular fluid.
The effective Casimir-like force between two intruders, induced 
by the fluctuations of the hydrodynamic fields, can change its sign 
when varying the control parameters: the volume fraction, the 
distance between the intruders, and the restitution coefficient. 
More interestingly, by inserting more intruders, we verify that 
the fluctuation-induced interaction is not pairwise additive. 
The simulation results are qualitatively consistent with the 
theoretical predictions based on mode coupling calculations. 
These results shed new light on the underlying mechanisms of 
collective behaviors in fluidized granular media.
\end{abstract}


\maketitle

Granular segregation has been extensively investigated during 
the last two decades aimed at revealing the underlying complex
dynamics \cite{1,2}. Besides the scientific
interest, understanding the mechanisms of segregation is of
essential importance in geophysical \cite{3} and
industrial \cite{4} processes. The behavior of granular 
mixtures, when mechanically agitated, depends on a long list 
of grain, container, and external driving properties 
\cite{2,5}. The control parameters can be
tuned so that the demixing is initiated, reversed, or prevented
\cite{5,6,7,8,9,10}.
While the phase behavior of these systems is still a matter of
debate, the nature of particle-particle interactions is known to
play a crucial role; two extreme limits can be distinguished: (i)
the fully fluidized regime where particles undergo only binary
collisions, and (ii) the lasting contacts regime where durable 
frictional contacts exist during a considerable part of the 
agitation cycle. While in the latter case the relevant 
processes are, e.g., reorganization, inertia, and convection 
\cite{11}, some studies reveal the existence 
of another mechanism in the fluidized regime: in the presence 
of intruder particles, the hydrodynamic fields are modified 
especially in the inner regions between intruders, leading to 
effective long-range interactions \cite{8,9,13,12}. Cattuto 
{\it et al.}\ \cite{12} found that 
a pair of intruder particles experience an effective force in 
a driven granular bed, originating from the modification of 
the pressure field fluctuations due to the boundary conditions 
imposed by the intruders. Such Casimir-like interactions are 
expected in thermal noisy environments confined by geometrical 
constraints \cite{14}. Most reports, so far, are about 
either binary mixtures \cite{5,6} or one or few 
intruder particles in a bed of smaller ones 
\cite{8,10,13}. An important question to address is 
how the collective behavior is influenced by the number 
and arrangement of the intruders.

\begin{figure}[b]
\centering
\includegraphics[scale=0.29,angle=-90]{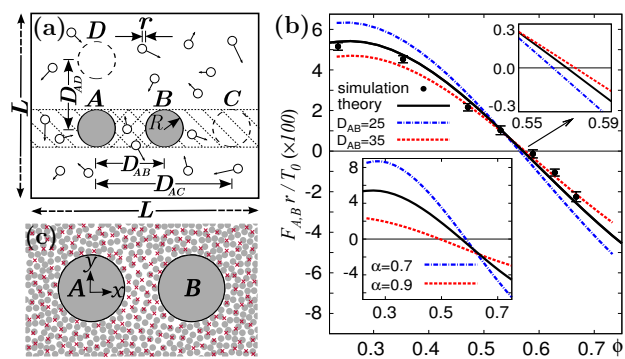}
\caption{(color online) (a) Sketch of the simulation cell. 
(b) $F_{_\text{A,B}}$ scaled by $T_{_0}/r$ vs $\phi$. 
Comparison is made with the solution of Eq.~(\ref{Equation-2}) 
for the reference system (solid line), as well as other 
values of the control parameters $D\!_{_{\text{AB}}}$ 
and $\alpha$ (dashed lines). (c) Typical snapshots of 
dense (gray circles) and dilute (red crosses) states 
with $\phi{=}0.66$ and $0.24$, respectively.}
\label{Fig1}
\end{figure}

In the present Letter, we study the effective 
interactions between immobile intruder particles immersed 
in a uniformly agitated granular fluid where all particles 
undergo inelastic binary collisions (Fig.~\ref{Fig1}). We 
show that the interaction between a pair of intruders 
exhibits a crossover from attraction to repulsion below 
a critical density, as predicted in \cite{12}. We 
here address the general conditions under which the 
transition happens, and present the phase diagram of 
the transition. Moreover, by comparing the behavior 
of two and multi intruder systems, we find that the 
fluctuation-induced force is not derived from a 
pair-potential; inserting a new intruder affects the 
previously existing interactions in a non-trivial way, 
depending on the relative positions of the intruders. Such 
a feature together with the possible sign change of the 
forces make the multi-body interactions more complicated 
and may lead to a variety of collective behaviors such 
as segregation, clustering, or pattern formation. 
Analytical calculations using the theory of randomly 
driven granular fluids \cite{15} confirm our 
findings.

{\it Simulation method ---} We consider a 2D granular 
fluid similar to the setup described in 
Refs.\cite{16,15,12} by means of 
molecular dynamics simulations. We have a reference 
system with two intruders A and B in which $L_{_0}/r{=}200$,
$R_{_0}/r{=}10$, and $D\!_{_{\text{AB},0}}/r{=}30$ [see 
Fig.~\ref{Fig1}(a)]. The reference volume fraction 
$\phi_{_0}$ is $0.66$, and the normal restitution coefficient 
$\alpha_{_0}$ is set to 0.8 for all collisions. Periodic 
boundary conditions are applied in both directions of the 
square-shaped cell to provide a spatially homogeneous state. 
The system is coupled to an external heat bath that uniformly 
transfers energy into the system; the acceleration of each 
particle $\bm a_i$ is perturbed instantaneously by a random 
amount $\bm \xi_i$ which can be considered as a Gaussian
white noise with zero mean and correlation $\langle \xi_{ia}(t)
\xi_{jb}(t^{\prime})\rangle{=}\xi^2 \delta_{ij} \delta_{ab}
\delta(t-t^{\prime})$, where $a$ and $b$ denote Cartesian
components of the vectors, and $\xi$ is the driving strength. 
The rate of the energy gain of a single particle averaged over 
the uncorrelated noise source is $\partial_t E {=} m\xi^2$
\cite{15}. Each particle also loses energy due to 
inelastic collisions at the mean-field rate of $\partial_t E 
{=} (\alpha^2\!-\!1)\omega T /2$ \cite{18,17}, 
where $T$ is the granular temperature and $\omega$ is 
the collision frequency given by the Enskog theory \cite{19}. 
Eventually, the system reaches a nonequilibrium stationary 
state by balancing the energy input and the dissipation.

\begin{figure}[t]
\centering
\includegraphics[scale=0.69,angle=0]{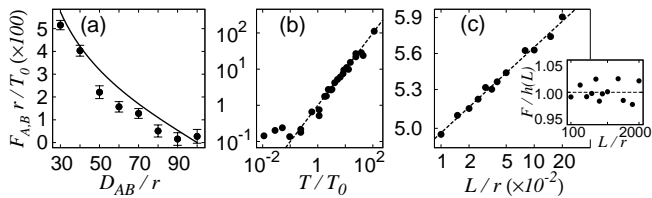}
\caption{(a) $F\!\!_{_\text{A,B}}$ vs $D\!_{_\text{AB}}$. 
The solid line is obtained via Eq.~(\ref{Equation-2}). (b) 
$F\!\!_{_\text{A,B}}$ vs $T$. The dashed line indicates a 
linear relation. Values of $F$ around $10^{-3} T\!_{_0}/r$ 
reflect the accuracy level of our calculation. (c) 
$F\!\!_{_\text{A,B}}$ vs $L$. The dashed line corresponds 
to ${h}(L)$. The inset shows that the deviation of the 
results from ${h}(L)$ has no systematic dependence on $L$. 
$\phi{=}0.24$ in all cases.}
\label{Fig2}
\end{figure}

{\it Effective two-body interactions ---} In the steady state 
we measure the total force exerted by the granular fluid 
on each intruder along the $x$ axis during the time interval 
$\tau$ ($\tau \!\!\sim\!\! 250$ collisions per particle). Due 
to the observed large fluctuations, the force is measured for 
more than $10^4$ consecutive time intervals $\tau$. The 
probability distribution of the data is well fitted by a 
Gaussian \cite{20} with the standard deviation 
$\sigma {=} 0.244 T\!_{_0}/r$ and the nonzero mean 
$F_{_0}{=}-0.023 T\!_{_0}/r$, where $T\!_{_0}$ is the 
mean-field approximation of the steady-state temperature 
deduced from the Enskog theory \cite{15}. Using a 
similar analysis along the $y$ axis, we obtain zero force 
within the accuracy of our measurements. $F_{_0}$ can be 
considered as the magnitude of the effective force 
$F\!\!_{_\text{A,B}}$ that the intruder B exerts on A, 
which is attractive in this case. We observe that, upon 
decreasing the volume fraction below a critical value 
$\phi_{_c} \!\!\sim\!\! 0.57$, the effective interaction 
$F\!\!_{_\text{A,B}}$ becomes repulsive [Fig.~\ref{Fig1}(b)], 
in agreement with the prediction of Ref.~\cite{12}. However, 
the transition is controlled not only by $\phi$, but also 
by $D\!_{_\text{AB}}$ and $\alpha$.  
One expects that, far from the transition region, increasing 
$D\!_{_\text{AB}}$ decreases the magnitude of $F\!\!_{_\text{A,B}}$ 
and it should eventually vanish at $D\!_{_\text{AB}}{=}L/2$ 
due to periodic boundary conditions, as confirmed by simulations 
[Fig.~\ref{Fig2}(a)]. In 
Fig.~\ref{Fig2}(b), by varying the 
driving strength $\xi$, it is shown that $F$ is proportional 
to the steady state temperature. Moreover, the results 
reveal the impact of dimensionality on the process: $F$ 
increases slightly with $L$ when we vary the system size 
while other parameter values are kept fixed. The simulation 
results, shown in Fig.~\ref{Fig2}(c), 
can be well fitted by a logarithmic growth $h(L) {=} a\ln(L) {+} b$ 
(dashed line). This is contrary to what happens in three 
dimensional systems, where the force is independent of $L$.

\begin{figure}[b]
\centering
\includegraphics[scale=0.31,angle=-90]{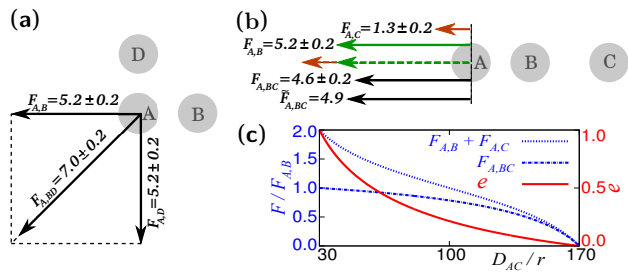}
\caption{(color online) Comparison between the binary and triple 
effective forces (scaled by $0.01 T\!_{_0}/r$) exerted on particle 
A in the presence of particles (a) B and D (b) B and C. (c) 
$e$, $F\!\!_{_\text{A,BC}}$ and $F\!\!_{_\text{A,B}} \!\!+\! 
F\!\!_{_\text{A,C}}$ versus $D\!_{_\text{AC}}$, the $x$ position 
of particle C.} 
\label{Fig3}
\end{figure}

\begin{figure*}[t]
\centering
\includegraphics[scale=0.98,angle=0]{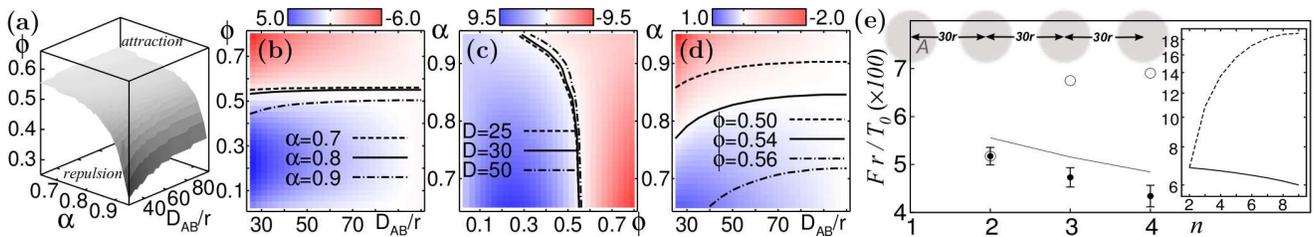}
\caption{(color online) (a) Schematic phase diagram of the 
transition in the ($\phi$, $D_\text{AB}$, $\alpha$) space. 
The surface corresponds to $F\!\!_{_\text{A,B}}{=}0$. (b-d) 
2D profiles of the phase diagram where the solid lines mark 
the interface position. The dashed lines correspond to the 
interface position for other values of the third control 
parameter. The color intensity reflects the magnitude of 
$F$ (scaled by $0.01 T\!_{_0}/r$), with blue (red) meaning 
repulsion (attraction). (e) $F$ on particle 
A vs $n$, the length of the chain. The results of the mode 
coupling, simulation, and pairwise summation of two-body 
forces are shown with solid line, full circles, and open 
circles, respectively. Inset: Mode coupling results (solid 
line) and their corresponding pairwise summations (dashed 
line) ($L{=}500r$).} 
\label{Fig4}
\end{figure*}

{\it Triple configurations and nonadditivity ---} Next we 
address the interesting case of triple systems, where the 
third intruder is located either on the $x$ or $y$ axis 
(Fig.~\ref{Fig3}). By choosing 
$D\!_{_\text{AD}}\!/\!r{=}30$ and $\phi{=} 0.24$, the 
effective force $\bm F\!\!_{_\text{A,BD}}$ exerted on particle A
in the triple configuration (A,B,D) is compared to $\bm
F\!\!_{_\text{A,B}}$ and $\bm F\!\!_{_\text{A,D}}$ obtained from
the binary systems (A,B) and (A,D), respectively. Note that the
simulation is performed anew for each set of intruders. Figure 
\ref{Fig3}(a) shows that $\bm F\!\!_{_\text{A,BD}}$ 
is nearly the vector sum of $\bm F\!\!_{_\text{A,B}}$ and $\bm
F\!\!_{_\text{A,D}}$. However, a comparison between the sets 
(A,B,C), (A,B), and (A,C) in Figure~\ref{Fig3}(b) 
(with $D\!_{_\text{AC}}\!/\!r{=}70$) reveals that the force is 
definitely not pairwise additive in this case; $F\!\!_{_\text{A,BC}}$ 
is even smaller than $F\!\!_{_\text{A,B}}$.

In order to understand the mechanism behind the long-range 
interactions and the transition, we draw attention to the fact 
that the fluctuating hydrodynamic fields, e.g.\ density [see 
Fig.~\ref{Fig1}(c)], are notably influenced by the geometric 
constraints, resulting in pressure imbalance around the 
intruders and effective interactions between them. To 
establish a quantitative connection between the 
effective force and the hydrodynamic fluctuations, 
we first employ mode coupling calculations \cite{15,21,12} to 
evaluate the two-body interactions. In the nonequilibrium steady 
state, the hydrodynamic fields ($p(\bm r)$,$T(\bm r)$,$n(\bm r)$) 
fluctuate around their stationary values ($p_s$,$T\!_s,n_s$). 
The average pressure fluctuation $p_{_f}\!(\bm r)$
in the presence of the boundary conditions imposed by intruders
behaves analogously to the Casimir effect, i.e., $p_{_f}\!(\bm r)$ 
in the hatched region of Fig.~\ref{Fig1}(a) differs from 
that of the cross-hatched region. Using the Verlet-Levesque 
equation of state for a hard disks system $p(n,T) {=} Tf(n)$ (with 
$f(n) {=}n(1+\phi^2/8)\!/\!(1-\phi)^2$, and $n$ the number density)
\cite{22}, we expand the pressure up to second order around
($n_s,T\!_s$), and take the statistical average over the random
noise source: $p_{_f}(\bm r){=} f^\prime(n)|_{_{n\!_s}}
\langle \delta n(\bm r) \delta T(\bm r) \rangle \!+\! \frac{1}{2}
T\!_s f^{\prime\prime}(n)|_{_{n\!_s}} \langle (\delta n(\bm r))^2
\rangle$. By Fourier transforming $\delta n(\bm r)$ and $\delta
T(\bm r)$ one obtains
\begin{equation}
p_{_f}\!(\bm r) {=} \int \!\! \left[ f^\prime(n)|_{_{n\!_s}}
S_\text{nT}(\bm k) \! + \! \frac{1}{2} T\!_s
f^{\prime\prime}(n)|_{_{n\!_s}} S_\text{nn}(\bm k)  \right] d\bm k,
\label{Equation-1}
\end{equation}
where the integral is taken over the $\bm k$ vectors allowed at
position $\bm r$ by the boundary conditions, and $S_\text{ab}(\bm
k)$ is the pair structure factor defined as $V^{-1} \langle \delta
a(\bm k) \delta b(-\bm k) \rangle$. The detailed description of
the structure factor calculations will be reported elsewhere (see
also Ref.\ \cite{15}). Here we denote the integrand of
Eq.~(\ref{Equation-1}) with $g(\bm k, \phi, \alpha)$ and compare
$p_{_f}\!(\bm r)$ for two surface points located on opposite
sides of intruder A with the same $y$ coordinates. The related
$\bm k$ vectors in the $x$ direction are confined to
$D\!_{_\text{in}}\!(y){=}D\!_{_\text{AB}}\!\!-\!
2\sqrt(R^2\!\!-\!\!y^2)$ and
$D\!_{_\text{out}}\!(y){=}L\!-\!D\!_{_\text{AB}}\!\!\!-\!\!2\sqrt(R^2\!\!-\!\!y^2)$
in the cross-hatched and hatched regions, respectively; Therefore
the pressure difference between these two points $\Delta
p(y){=}p^\text{(in)}_{_f}\!-\!p^\text{(out)}_{_f}$ has $y$
dependence. By integrating over $y$, we arrive at the average
pressure difference between the gap and outside region:
\begin{equation}
\Delta p{=} \!\!\!\int_{-R}^R \!\!\!\!\!dy
\!\!\left[\! \int_{2\pi\!/\!D\!_{_\text{in}}\!\!(y)}^{2\pi\!/\!r^*}
\!\!\!\!\!\!\!\!\!\!\! dk_{_x} \!\!\!-\!\!\!
\int_{2\pi\!/\!D\!_{_\text{out}}\!(y)}^{2\pi\!/\!r^*}
\!\!\!\!\!\!\!\!\! dk_{_x} \right] \!\!\int_{2\pi\!/\!L}^{2\pi\!/\!r^*}
\!\!\!\!\!\!\! dk_{_y} \!\!\frac{g(k_x,k_y,\phi,\alpha)}{2R} .
\label{Equation-2}
\end{equation}
To ensure that the hydrodynamic description is valid, the
integrals are restricted to the long wavelength range
$r^*{=}\text{max}(2r,l^*)$ (with $l^*$ being the mean 
free path) and only small inelasticities are considered. 
Moreover, $D\!_{_\text{AB}}$ is always chosen 
large enough ($3R {\leq} D\!_{_\text{AB}}$) so that the 
short-range depletion forces \cite{23} do not play a 
role. The effective force $F\!_{_\text{AB}}\!$ is calculated 
via Eq.~(\ref{Equation-2}) for different values of $\phi$ or 
$D\!_{_\text{AB}}$ and compared to the simulation results in 
Figs.~\ref{Fig1}(b) and \ref{Fig2}(a). 
The forces are of the same order of magnitude as those obtained 
from the simulations. The deviations can be attributed to the 
fact that the hydrodynamic fluctuations are correlated in the 
gap and outside regions. The correction due to this effect is
proportional to $\partial^2 g/\partial \phi^2$ which always has a
sign opposite to that of $g$, thus, Eq.~(\ref{Equation-2})
overestimates the magnitude of the force. Using the mode coupling
calculations, it is shown in Fig.~\ref{Fig1}(b) that the
transition point is sensitive to the choice of $D\!_{_\text{AB}}$
and $\alpha$. The set of control parameters for sign 
switching of the Casimir force crucially depends on the 
physics of the system (see e.g.\ \cite{24}). Figure 
\ref{Fig4} summarizes the calculations in a 
\emph{phase diagram} in the ($\phi$, $\alpha$, $D\!_{_\text{AB}}$) 
space, which appears to be in remarkable accord with the dynamical 
model. The phase diagram is not 
influenced by the choice of the steady-state temperature 
$T\!_s$, while the magnitude of the force grows linearly 
with $T\!_s$ as expected for Casimir forces in thermal 
fluctuating media \cite{14}. Regarding the fact, that 
the leading term of $g(\bm k, \phi, \alpha)$ at small $k$ 
is proportional to $1/k^2$ \cite{15}, one also finds 
from Eq.~(\ref{Equation-1}) that $p_{_f}$, and therefore the 
force, in the thermodynamic limit behaves as $1/r$ in 3D 
while diverges logarithmically as $\text{ln}(L/r)$ in 2D.

{\it Multi-body effects ---} In the triple system of 
Fig.~\ref{Fig3}(a), loosely speaking, because 
of the independence of $k$ vectors in $x$ and $y$ directions 
one expects that the vector sum holds. This is in agreement 
with the simulation results in Fig.~\ref{Fig3}(a), 
neglecting the deviation due to hydrodynamic correlations. 
In the configuration of Fig.~\ref{Fig3}(b), 
the effective force exerted on particle A results from the 
difference between the range of available $k$ modes on its 
left and right sides. In the presence of particle C, the 
range of allowed $k$ modes decreases on the left side due 
to periodic boundary conditions, which 
causes a lowering of pressure difference between both sides 
of A. The effective force $F\!\!_{_\text{A,BC}}$ is thus 
smaller, compared to the binary interaction 
$F\!\!_{_\text{A,B}}$. The calculated force on particle A using
Eq.~(\ref{Equation-2}) is shown with 
$\widetilde{F}\!\!_{_\text{A,BC}}$ in
Fig.~\ref{Fig3}(b); the agreement is satisfactory. 
We introduce a measure $e{=}|\bm F\!\!_{_\text{A,B}} \!\!+\!\!
\bm F\!\!_{_\text{A,C}}\!\!-\!\!\bm F\!\!_{_\text{A,BC}}|/| \bm
F\!\!_{_\text{A,B}}|$ for quantifying the deviation from the case
that the interactions are derived from a pair-potential. Figure
\ref{Fig3}(c) shows how $e$, obtained from the mode
coupling calculations, behaves when the position of particle C is
varied on the $x$ axis. Figure \ref{Fig4}(e) indicates that our analytical 
approach also provides reasonable estimates of the effective force 
acting on particle A in a chain configuration, while the deviation 
from pairwise summation of two-body forces grows with the number 
of intruders, $n$.

\begin{figure}[t]
\centering
\includegraphics[scale=0.42,angle=0]{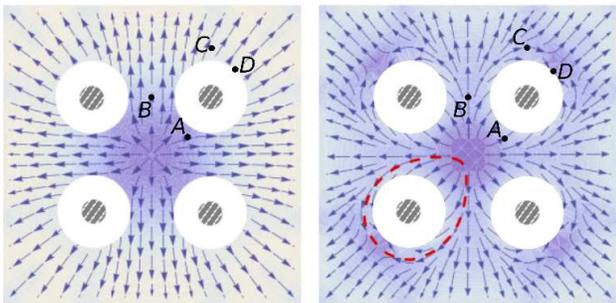}
\caption{The trajectories of a test intruder subject to the 
effective force field in a fixed square configuration of 
intruders, obtained from the mode coupling calculations for 
$\phi {=} 0.24$ (left), and $0.54$ (right). Short range 
interactions are excluded (white zones). Lighter colors 
mean stronger forces. The dashed line in the right figure 
indicates the border of an attractive subdomain, where the 
trajectories of the test particle end up on the fixed intruder.} 
\label{Fig5}
\end{figure}

\begin{table}[b]
\caption{Comparison between the theoretical (T) and simulation 
(S) results for the selected points in Figs.~\ref{Fig5}(left) 
and \ref{Fig5}(right). Same force scales as Fig.~\ref{Fig3}. 
The maximum error bars for the simulated $F$ and $\theta$ are 
$\pm 0.1$ and $\pm 7^{\circ}$, respectively.}
\begin{center}
\begin{tabular}{ccccccccc}
&$|\vec F|_{(A)}$&$|\vec F|_{(B)}$&$|\vec F|_{(C)}$&$|\vec F|_{(D)}$
&$\theta_{(A)}$&$\theta_{(B)}$&$\theta_{(C)}$&$\theta_{(D)}$\\
\hline
left-S&$1.0$& $2.2$&$3.4$&$3.7$&$13^{\circ}$&$92^{\circ}$
&$51^{\circ}$&$44^{\circ}$\\
left-T&$1.17$& $2.52$&$3.97$&$4.32$&$7^{\circ}$&$90^{\circ}$
&$60^{\circ}$&$45^{\circ}$\\
right-S&$0.2$&$0.2$&$0.2$&$0.0$&$19^{\circ}$&$88^{\circ}$
&$-32^{\circ}$&$-$\\  
right-T&$0.24$&$0.26$&$0.25$&$0.02$&$28^{\circ}$&$90^{\circ}$
&$-24^{\circ}$&$225^{\circ}$
\label{Table1}
\end{tabular}
\end{center}
\end{table}
Finally, we investigate more complicated geometries by 
calculating the effective force field of a fixed square 
structure acting on a test intruder particle. To 
elucidate the impact of sign switching on the force 
pattern we compare an intermediate density (around the 
transition zone) with a low density case. Figure 
\ref{Fig5} shows that the patterns are clearly 
different e.g.\ in terms of the number of equilibrium 
points. There exist also attractive subdomains of 
length scale $\ell$ around the fixed intruders in the right 
figure (e.g. the domain surrounded by the dashed 
line). Depending on the choice of control 
parameters, $\ell$ ranges between $0$ (entirely 
repulsive patterns at low densities, as shown in the left 
figure) and $L$ (entirely attractive patterns at high 
densities, not shown) leading to different types of 
collective behavior such as segregation, clustering, 
and pattern formation. Putting aside the pairwise 
summations, we compare the mode coupling predictions 
with simulation results at four selected points in 
Fig~\ref{Fig5}. The predicted forces are 
of the same order of magnitude as the simulation 
results (see table~\ref{Table1}), however, the 
errors of the force size $|F|$ and its direction 
$\theta$ reach up to $30\%$ and $15^{\circ}$, 
respectively. The differences are smaller at large 
distances and also low densities. Indeed, the 
hydrodynamic correlations become more important 
in multi-body cases and, hence, one should include 
triplet or higher order structure factors \cite{25} 
in mode coupling calculations to properly take the 
multi-body effects into account.

In conclusion, we focus on the problem of long-range
fluctuation-induced forces between intruder particles 
immersed in an agitated fluid bed. The sign of the 
force can be reversed by tuning the control parameters. 
Furthermore, the multi-body interactions do not follow from 
two-body force descriptions. A newly inserted 
intruder, depending on its position, may affect the pressure 
balance around the other intruders. This suggests that the 
effective force is not derived from a pair-potential in 
agreement with our simulation results. Our findings 
represent a step forward in understanding the origin of 
collective behaviors in fluidized granular mixtures.

We would like to thank I. Goldhirsch for helpful discussions and 
J. T\"or\"ok and B. Farnudi for comments on the manuscript. 
Computing time was provided by John-von-Neumann Institute of 
Computing (NIC) in J\"ulich.

\end{document}